\documentclass{jnmp}

\usepackage{amsmath}
\JNMPnumberwithin{equation}{section}

\newtheorem{theorem}{Theorem}

\theoremstyle{definition}

\begin{document}

%  Headings
%
\renewcommand{\evenhead}{M Czachor}
\renewcommand{\oddhead}{Reducible representations of CAR and CCR}

%  Titlepage
%
\thispagestyle{empty}

\FirstPageHead{*}{*}{20**}{\pageref{firstpage}--\pageref{lastpage}}{Article}
%  Parameters: Volume, number, year, page range, paper type
%  'Article' could be changed to 'Letter' or 'Review Article'

\copyrightnote{2003}{M Czachor}

\Name{Reducible representations of CAR and CCR with possible applications to field quantization}

\label{firstpage}

\Author{Marek CZACHOR}

\Address{
Departement Natuurkunde, Universiteit Antwerpen UIA,
B2610 Antwerpen, Belgium\\
and\\
Katedra Fizyki Teoretycznej i Metod Matematycznych\\
Politechnika Gda\'nska, 80-952 Gda\'nsk, Poland\\
E-mail: mczachor@pg.gda.pl}

\Date{Received Month *, 200*; Revised Month *, 200*; 
Accepted Month *, 200*}

\begin{abstract}
\noindent
Reducible representations of CAR and CCR are applied to second quantization of Dirac and Maxwell fields. The resulting field operators are indeed operators and not operator-valued distributions. Examples show that the formalism may lead to a finite quantum field theory.
\end{abstract}

%  The paper
%
\section{Reducible representation of CAR}

The main objective of the paper is to discuss a reducible representation of canonical anticommutation relations (CAR) which generalizes to Dirac electrons the construction previously employed in \cite{I,II,III} to electromagnetic fields. 
The case of canonical commutation relations (CCR) is briefly discussed in the last section. 

Beginning with the CAR operators $b_\pm$,  $d_\pm$ of an {\it irreducible\/} representation  we introduce the
following four operators 
\begin{equation}
b(\vec p,\pm)
=
|\vec p\rangle\langle \vec p|\otimes b_\pm=c_1(\vec p,\pm),\quad
d(\vec p,\pm)
=
|\vec p\rangle\langle \vec p|\otimes d_\pm=c_2(\vec p,\pm)
\end{equation}
satisfying a reducible representation of CAR. 
The momentum eigenvectors are normalized by 
$%\begin{equation}
\langle \vec p|\vec p\,'\rangle
=
\delta_{\Gamma_m}(\vec p,\vec p\,')
=
(2\pi)^3 2\sqrt{\vec p^2+m^2}\delta^{(3)}(\vec p-\vec p\,')
$.%\end{equation}
The reducible representation of CAR can be written in a compact form as 
\begin{equation}
\big\{c_j(\vec p,s),c_{j'}(\vec p\,',s')^{\dag}\big\}
=
\delta_{jj'}\delta_{ss'}\delta_{\Gamma_m}(\vec p,\vec p\,')
|\vec p\rangle\langle \vec p|\otimes 1
=
\delta_{jj'}\delta_{ss'}\delta_{\Gamma_m}(\vec p,\vec p\,')
I_{\vec p},\label{nCAR}
\end{equation}
the remaining anti-commutators vanishing.
The identity 1 in (\ref{nCAR}) is the one
occuring in the CAR relations 
$\{b_\pm,b_{\pm}^{\dag}\}=\{d_\pm,d_{\pm}^{\dag}\}=1$
and the RHS of (\ref{nCAR}) is in the center of the CAR algebra. 
Similarly to \cite{I,II,III}  we 
have introduced the operator 
$
I_{\vec p}=|\vec p\rangle\langle \vec p|\otimes 1
$
satisfying the resolution of unity 
$
\int d \Gamma_m(\vec p)
I_{\vec p}=
\int d \Gamma_m(\vec p)
|\vec p\rangle\langle \vec p|\otimes 1
=I.
$
We define the single-oscillator Dirac field operator by 
\begin{gather}
\Psi(x)=\sum_s\int d \Gamma_m(\vec p)\Big(
u(\vec p,s) b(\vec p,s)e^{-ip\cdot x}
+
v(\vec p,s) d(\vec p,s)^{\dag}e^{ip\cdot x}
\Big).
\end{gather}
In order to perform the second step of quantization we introduce 
$
I_0=\int d \Gamma_m(\vec p)|\vec p\rangle\langle \vec p|\otimes 1_0
$
where $1_0$ is a Hermitian operator satisfying $1_0^2=1$, 
$\{b_\pm,1_0\}=\{d_\pm,1_0\}=0$. 
A fermionic $N$-oscillator Jordan-Wigner-type extension is defined by 
\begin{gather}
\underline{b}(\vec p,s)
=
\frac{1}{\sqrt{N}}
\sum_{n=1}^{N}
\overbrace{
\underbrace{I_0\otimes\dots\otimes I_0}
_{n-1}  \otimes\, b(\vec p,s)\otimes I \otimes \dots
\otimes I}^{N}
=
\underline{ c}_1(\vec p,s),\\
\underline{ d}(\vec p,s)
=
\frac{1}{\sqrt{N}}
\sum_{n=1}^{N}
\overbrace{
\underbrace{I_0\otimes\dots\otimes I_0}
_{n-1}  \otimes\, d(\vec p,s)\otimes I \otimes \dots
\otimes I}^{N}
=
\underline{ c}_2(\vec p,s),\\
\underline{\Psi}(x)
=
\sum_s\int d \Gamma_m(\vec p)\Big(
u(\vec p,s) \underline{ b}(\vec p,s)e^{-ip\cdot x}
+
v(\vec p,s) \underline{ d}(\vec p,s)^{\dag}e^{ip\cdot x}
\Big).
\end{gather}
The reducible representation of CAR reads
\begin{gather}
\big\{\underline{ c}_j(\vec p,s),\underline{ c}_{j'}(\vec p\,',s')^{\dag}\big\}
=
\delta_{jj'}\delta_{ss'}\delta_{\Gamma_m}(\vec p,\vec p\,')
\underline{ I}_{\vec p},\\
\underline{ I}_{\vec p}
=\frac{1}{N}
\sum_{n=1}^{N}
\overbrace{
\underbrace{I\otimes\dots\otimes I}
_{n-1}  \otimes I_{\vec p}\otimes I \otimes \dots
\otimes I}^{N},\quad
\int d \Gamma_m(\vec p)
\underline{I}_{\vec p}=\underline{I}.
\label{uu nCAR}
\end{gather}
In order to verify that 
$\underline{\Psi}(x)$ is an operator it is sufficient to check this property for 
$\Psi(x)$. 
The choice of the representation implies that 
\begin{gather}
\Psi(x)
=
\sum_s\Big(
u(\hat{\vec p},s) e^{-i\hat p\cdot x}\otimes b_s
+
v(\hat{\vec p},s) e^{i\hat p\cdot x} \otimes d_{s}^{\dag}
\Big)\nonumber
\end{gather}
where 
$
\hat{p}_a=\sum_s\int d \Gamma_m(\vec p)
p_a|\vec p\rangle\langle \vec p|
$
is the spectral representation of the unbounded operator, 
and $u(\hat{\vec p},s)$, $v(\hat{\vec p},s)$
are functions of the operator $\hat{\vec p}$ in the sense of spectral theory.
All these objects are well defined and there is no problem with products of fields taken at the same point $x$ of the configuration space. 
The difference between fields taken in our reducible representation and those arising from the standard irreducible one is analogous to this between the unitary operator $e^{-i\hat p\cdot x}=
\int d \Gamma_m(\vec p)e^{-ip\cdot x}|\vec p\rangle\langle \vec p|$ 
and the distribution $\int d \Gamma_m(\vec p)e^{-ip\cdot x}$. Unitary representation of the Poincar\'e group
\begin{equation}
U_{\Lambda,y}^{-1}\underline{\Psi}_\alpha(x)U_{\Lambda,y}=
S_{\alpha}{^\beta} \underline{\Psi}_\beta\big(\Lambda^{-1}(x-y)\big)
\end{equation}
is explicitly constructed in \cite{IV}. Our representation of CAR written in terms of Dirac fields reads (the superscripts $(+)$ or $(-)$ denote the parts of field operators containing only creation or annihilation operators, respectively) 
\begin{gather}
\underline{ S}^{(\pm)}_{\alpha\beta}(x-y)
\stackrel{\rm def}{=}
i\{\underline{\Psi}^{(\mp)}_\alpha(x),\underline{\bar\Psi}^{(\pm)}_\beta(y)\}
=
\frac{1}{N}
\sum_{n=1}^{N}
\overbrace{
\underbrace{I\otimes\dots\otimes I}
_{n-1}  \otimes 
S^{(\pm)}_{\alpha\beta}(x-y)
\otimes I \otimes \dots
\otimes I}^{N}
\nonumber\\
S^{(\pm)}_{\alpha\beta}(x-y)
\stackrel{\rm def}{=}
i\{\Psi^{(\mp)}_\alpha(x),\bar\Psi^{(\pm)}_\beta(y)\}
=
i(\gamma \cdot \hat p\pm m)_{\alpha\beta}e^{\mp i\hat p\cdot(x-y)}\otimes 1.
\end{gather} 
Neither anti-commutators nor products of fields evaluated at 
$x=y$ lead to any difficulty. The essence of the new representation is in the replacement of ordinary integrals over momenta by spectral integrals (e.g.  regularization of zero-point energy in a finite volume would take the form $\sum n \to \sum n|n\rangle\langle n|$).   
The limit $N\to\infty$ provides a correspondence principle analogous to $\hbar\to 0$ or $c\to\infty$. For physical aspects of the modification cf. \cite{I,II}.

\section{Vacuum and multi-electron states}

The vacuum consists of a Hilbert space of all the states which are annihilated by all annihilation operators. We begin with a ``single-oscillator vacuum" 
\begin{gather} 
|O\rangle
=
\int d \Gamma_m(\vec p)O(\vec p)|\vec p\rangle\otimes|0\rangle,
\end{gather} 
where $b_s|0\rangle=d_s|0\rangle=0$, i.e. $|0\rangle$ is the vacuum of the irreducible representation of CAR. The multi-oscillator vacuum is defined at 
the $N$-oscillator level as the tensor product of one-oscillator vacua
$
|\underline{O}\rangle
=
|O\rangle\otimes\dots\otimes|O\rangle.
$
As expected
$
\underline{b}(\vec p,s)|\underline{O}\rangle=
\underline{d}(\vec p,s)|\underline{O}\rangle=0.
$
Let us stress that the vacuum {\it space\/} as a whole is Poincar\'e invariant, whereas a concrete vector $|\underline{O}\rangle$ is only covariant: The wave function transforms as a mass-$m$ scalar field which carries the unitary representation $O(\vec p)\mapsto e^{-2i p\cdot y}O(\overrightarrow {\Lambda^{-1}p})$ \cite{II,IV}; the exponent comes from the vacuum part of 4-momentum and can be removed by a unitary transformation. The vacuum space plays a role of a base space in a bundle whose fibers are Fock spaces generated in the usual way in terms of the representation of CAR. The structure of the fibers is characterized by a theorem on the thermodynamic limit $N\to\infty$ (the proof is exactly analogous to the bosonic case explained in detail in \cite{II,III}). 
Let
$
\underline{c}_j(f)
=
\sum_{s}\int d \Gamma_m(\vec p) \overline{f(\vec p,s)}
\underline{c}_j(\vec p,s).
$
The scalar product of two unnormalized one-electron states is 
\begin{gather} 
\langle \underline{O}|\underline{c}_j(f) \underline{c}_{j'}(g)^{\dag}
|\underline{O}\rangle
=
\delta_{jj'}\sum_{s}\int d \Gamma_m(\vec p)Z(\vec p)
\overline{f(\vec p,s)}g(\vec p,s)
\stackrel{\rm def}{=}
\delta_{jj'}\langle f|g\rangle_Z\label{<of|og>}
\end{gather} 
where $Z(\vec p)=|O(\vec p)|^2$. 
The scalar product $\langle f|g\rangle_Z$ reappears in the thermodynamic limit $N\to\infty$ for arbitrary multi-electron states.
\begin{theorem}
Let $|\underline{O}\rangle
=
\underbrace{|O\rangle\otimes\dots\otimes|O\rangle}_N$. Then
\begin{gather} 
\lim_{N\to\infty}
\langle \underline{O}|\underline{c}_j(f_1)\dots \underline{c}_j(f_M)
\underline{c}_j(g_1)^{\dag}\dots
\underline{c}_j(g_M)^{\dag}|\underline{O}\rangle
=
\sum_{\sigma}\delta_\sigma
\langle f_1|g_{\sigma(1)}\rangle_Z
\dots 
\langle f_M|g_{\sigma(M)}\rangle_Z\nonumber\\
=
\sum_{\sigma}\delta_\sigma\sum_{s_1\dots s_M}
\int d\Gamma_m(\vec p_1)\dots
\int d\Gamma_m(\vec p_M)
\nonumber\\
\phantom{=}
\times
Z(\vec p_1)\dots Z(\vec p_M)\overline{f_1(\vec p_1,s_1)}\dots 
\overline{f_m(\vec p_M,s_M)}
g_{\sigma(1)}(\vec p_1,s_1)
\dots 
g_{\sigma(m)}(\vec p_M,s_M)\nonumber
\end{gather} 
where $\delta_\sigma$ is the sign of the permutation $\sigma$.
\end{theorem}
Let us note that the weights 
$Z(\vec p)=\langle \underline{O}|\underline{I}_{\vec p}|\underline{O}\rangle$ occur here automatically and regularize ultraviolet divergences since 
$\int d \Gamma_m(\vec p)Z(\vec p)=
\langle \underline{O}|\int d \Gamma_m(\vec p)
\underline{I}_{\vec p}|\underline{O}\rangle=
\langle \underline{O}|\underline{I}|\underline{O}\rangle=1$
implies that $Z(\vec p)$ decays at infinity. Furthermore, the Poincar\'e transformation $Z(\vec p)\mapsto Z(\overrightarrow {\Lambda^{-1}p})$ implies that 
$Z=\max_{\vec p}\{Z(\vec p)\}$ is a nonvanishing invariant. This allows us to introduce the cut-off functions $\chi(\vec p)=Z(\vec p)/Z$, $0\leq \chi(\vec p)\leq 1$. Quantum electrodynamic regime is defined as a condition on supports of wave packets: $f(\vec p,s)\chi(\vec p)=f(\vec p,s)$. For any two such wave packets one finds 
$\langle f|g\rangle_Z=Z\langle f|g\rangle$ and the thermodynamic limit simply redefines a bare charge by a multiple of the invariant $Z$, as we shall see explicitly in the next section. 

\section{Interaction Hamiltonian}

For Dirac electrons interacting with classical electromagnetic fields the departure point is the interaction-picture Hamiltonian 
\begin{equation}
H(x_0)
=
e_0 \int d^3x 
\underline{\bar\Psi}(x)\gamma_a\underline{\Psi}(x)
A^a(x)
\end{equation}
Let us note that
$%\begin{equation}
H(x_0)
=
-i(2\pi)^3e_0 \underline{S}^{(-)}_{\alpha\beta}(0)
\gamma_a^{\alpha\beta}\tilde A^a(x_0,0)
+
:H(x_0):\,
$%\end{equation} 
where $\tilde A^a(x_0,\vec k)$ is the 3-dimensional Fourier transform. The difference between $:H(x_0):$ and $H(x_0)$ is a well behaved element of the center of CAR if the Fourier transform is not singular at $\vec k=0$. This shows that the Coulomb field leads to an infrared vacuum divergence. However, if one restricts $A_a(x)$ to solutions of free Maxwell equations then $\tilde A^a(x_0,0)=0$ since the origin of the light cone is excluded. We shall return to this question in the next section.

As a first purely fermionic modification consider the vacuum average
\begin{equation}
\langle\underline{O}|\underline{S}^{(\pm)}(x)|\underline{O}\rangle
=
\int d\Gamma_m(\vec p)
\langle \underline{O}|\underline{I}_{\vec p}|\underline{O}\rangle
(\gamma \cdot p\pm m)e^{\mp ip\cdot x}
=
Z
\int d\Gamma_m(\vec p)\chi(\vec p)
(\gamma \cdot p\pm m)e^{\mp ip\cdot x}.\nonumber
\end{equation}
This is the usual-looking expression but with the cut-off function 
$\chi(\vec p)$ and the ``renormalization constant" $Z$ automatically built-in. 
The regularization by $\chi(\vec p)$ is a straightforward consequence of reducibility: For irreducible representations one finds $\langle \underline{O}|\underline{I}_{\vec p}|\underline{O}\rangle=1$.

A first-order correction in a $S$-matrix element evaluated between two single-electron wave packets reads 
\begin{gather}
\langle\phi|S^{(1)}|\psi\rangle
=
-ie_0
\sum_{s,s'}
\int d\Gamma_m(\vec p)\int d\Gamma_m(\vec p\,')
\\
\phantom{\langle{\rm out}|S^{(1)}|{\rm in}\rangle
=}\times
\underbrace{
\langle \underline{O}|\underline{I}_{p\,'}
\underline{I}_p|\underline{O}\rangle}_{\rm modification}
\overline{\phi(\vec p\,',s')}
\psi(\vec p,s)\bar u(\vec p\,',s')\gamma_a u(\vec p,s)
\int d^4x
e^{i(p\,'-p)\cdot x} A^a(x).\nonumber
\end{gather}
The underbraced expression, equal to 1 in the standard irreducible representation, is the only modification we encounter. Explicitly 
\begin{equation}
\langle \underline{O}|\underline{I}_{p\,'}
\underline{I}_p|\underline{O}\rangle
=
Z\frac{1}{N}\delta_{\Gamma_m}(\vec p,\vec p\,')
\chi(\vec p)
+
Z^2\Big(1-\frac{1}{N}\Big)
\chi(\vec p)\chi(\vec p\,').\label{1/N'}
\end{equation}
Keeping in mind that one power of $Z$ gets absorbed into normalization of single-electron states, assuming $|\int d^4x A^a(x)|<\infty$ we find, in the thermodynamic limit $N\to\infty$ and 
for wave packets satisfying the QED regime condition, that 
$\langle\phi|S^{(1)}|\psi\rangle=Z\langle\phi|S^{(1)}|\psi\rangle_{\rm standard}$. For finite $N$ there are additional corrections arising from the term proportional to $1/N$. 

As a next exercise consider the vacuum polarization tensor which appears in second-order calculation of vacuum polarization. The standard expression is here replaced by 
\begin{gather}
{\rm Tr\,}
\Big(
\langle\underline{O}|\underline{S}^{(-)}(x_2-x_1)
\gamma^a
\underline{S}^{(+)}(x_1-x_2)\gamma^b|\underline{O}\rangle
\Big)\nonumber\\
\phantom{{\rm Tr\,}}
=
-
\frac{Z}{N}
\int d\Gamma_m(\vec p)
\chi(\vec p)e^{2i{p}\cdot (x_2-x_1)}
{\rm Tr\,}
\Big((\gamma\cdot p-m)\gamma^a (\gamma\cdot p+m)\gamma^b\Big)
\nonumber\\
\phantom{{\rm Tr\,}=}
-Z^2
\Big(1-\frac{1}{N}\Big)
\int d\Gamma_m(\vec p) \int d\Gamma_m(\vec p\,')\nonumber\\
\phantom{{\rm Tr\,}==}\times
\chi(\vec p)\chi(\vec p\,')e^{i(p+p')\cdot (x_2-x_1)}
{\rm Tr\,}
\Big((\gamma\cdot p-m)\gamma^a (\gamma\cdot p'+m)\gamma^b\Big).
\nonumber
\end{gather}
The expression which survives the thermodynamic limit is a regularized version of the standard formula, multiplied by $Z^2$.

\section{Reducible representation of CCR}

The automatic appearance of the cut-off functions $\chi(\vec p)=|O(\vec p)|^2/Z$ was a consequence of replacing at the RHS of CAR the identity operator by 
$\underline{I}_{\vec p}$. The claim is that an analogous effect occurs for reducibly quantized electromagnetic fields \cite{I,II,III}. 

The strategy is similar to the CAR case. One starts with 
\begin{equation}
a(\vec k,\pm)
=
|\vec k\rangle\langle \vec k|\otimes a_\pm
\end{equation}
where $[a_s,a^{\dag}_{s'}]=\delta_{ss'}1$ is an irreducible representation of CCR and then performs a bosonic $\tilde N$-particle extension $a(\vec k,\pm)\mapsto 
\underline{a}(\vec k,\pm)$. The parameters $N$ and $\tilde N$ representing the numbers of fermionic and bosonic oscillators are in principle unrelated. A theorem analogous to Theorem~1 holds for the $\tilde N\to\infty$ limit. Coherent states of light are defined in the standard way in terms of a displacement operator 
\begin{gather}
\underline{\cal D}(f) = e^{\underline{a}(f)^{\dag}
-\underline{a}(f)},\quad
\underline{a}(f)=
\sum_{s}\int d \Gamma_0(\vec k) \overline{f(\vec k,s)}
\underline{a}(\vec k,s),\\
\underline{\cal D}(f)^{\dag}
\underline{a}(\vec k,\pm)
\underline{\cal D}(f)
=
\underline{a}(\vec k,\pm)+f(\vec k,\pm)\underline{I}_{\vec k}.\label{D}
\end{gather}
Let us note the important difference with respect to the irreducible representation: The shift in (\ref{D}) is proportional to $\underline{I}_{\vec k}$ and not to the identity operator. 
For $\tilde N\to\infty$ the statistics of excitations of a coherent state is Poissonian, as it should be for physical reasons. 
Now assume we have a classical transverse current $J_a(x)$ and the interaction-picture Hamiltonian 
\begin{equation}
H(x_0)
=
\int d^3x 
J_a(x)
\underline{A}^a(x)\label{HJ}
\end{equation}
where 
$\underline{A}^a(x)$ is a reducibly quantized vector potential in a Lorenz gauge. 
A simple calculation shows that the $S$ matrix, up to a unitary operator which belongs to the center of CCR, is given by the displacement operator and 
\begin{gather}
\underline{a}(\vec k,\pm)_{\rm out}
=
S^{\dag}
\underline{a}(\vec k,\pm)_{\rm in}
S
=
\underline{a}(\vec k,\pm)_{\rm in}+j(\vec k,\pm)\underline{I}_{\vec k}
=
\underline{\cal D}(j)^{\dag}
\underline{a}(\vec k,\pm)_{\rm in}
\underline{\cal D}(j)
\end{gather}
where $j(\vec k,\pm)$ are amplitudes of the two transverse components of the 4-dimensional Fourier transform of $J_a(x)$, restricted to the light cone, i.e. 
the same functions one finds in the standard formalism. But there is also a difference: The irreducible formalism would have produced 
$j(\vec k,\pm)\underline{I}$ and not
$j(\vec k,\pm)\underline{I}_{\vec k}$. 

The vacuum space of the representation is constructed in analogy to the CAR case. One begins with
$
|\tilde O\rangle
=
\int d \Gamma_0(\vec k)\tilde O(\vec k)|\vec k\rangle\otimes|0\rangle,
$
where $a_s|0\rangle=0$, and 
the $\tilde N$-oscillator extension is 
$
|\underline{\tilde O}\rangle
=
|\tilde O\rangle\otimes\dots\otimes|\tilde O\rangle.
$
The ``renormalization constant" $\tilde Z=\max_{\vec k}\{|\tilde O(\vec k)|^2\}$ is a nonvanishing Poincar\'e invariant and thus the cut-off function 
$\tilde \chi(\vec p)=|\tilde O(\vec p)|^2/\tilde Z$ is well defined. What is important, the field $\tilde O(\vec k)$ is now massless and therefore vanishes not only at infinity but also at $\vec k=0$. 

Having the ``out" fields we can compute their (``in"-)vacuum averages. The result is analogous to the standard one, but now the Fourier transform of the field involves the amplitude
$%\begin{equation}
f(\vec k,\pm)
=
\langle \underline{\tilde O}|\underline{I}_{\vec k}|\underline{\tilde O}\rangle
j(\vec k,\pm)
=|\tilde O(\vec k)|^2j(\vec k,\pm)
=\tilde Z\tilde \chi(\vec k)j(\vec k,\pm)
$ 
and not $f(\vec k,\pm)=j(\vec k,\pm)$ which would have occured in the irreducible formalism. The difference is subtle but of crucial importance since the property 
$\tilde\chi(0)=0$ can regularize the infrared divergence. In the irreducible case the radiation field produced by an accelerated pointlike charge involves the amplitude 
$j(\vec k,\pm)$ which instead of vanishing blows up at $\vec k=0$. In the reducible case we obtain a regularization which eliminates the infrared divergence if the vacuum 
$\tilde O(\vec k)$ is correctly chosen. The same concerns the average number of photons of the radiated field \cite{II,III}. Let us mention that the operators 
$\underline{I}_{\vec k}$ appear also in reducibly quantized solutions of Maxwell equations with classical currents and thus regularize classical divergences. 

To conclude, the reducible representations seem to produce the cut-off functions in exactly those places one expects them to occur. This is a consequence of the RHS of CAR and CCR where instead of identities one finds $\underline{I}_{\vec p}$ and $\underline{I}_{\vec k}$. At the level of amplitudes and in thermodynamic limits one finds the effective rule $\underline{I}_{\vec p}\to Z \chi(\vec p)$, 
$\underline{I}_{\vec k}\to \tilde Z \tilde \chi(\vec k)$. The analogy to  renormalized fields\footnote{I am indebted to prof. H. Grosse for drawing my attention to this point during our discussion in Bia{\l}owie\.za}
whose CAR and CCR relations involve at RHS the renormalization constants $Z_2$ and $Z_3$ supplemented by cut-offs is probably not accidental. An argument in favor of our approach is that the RHS of CAR and CCR are Poincar\'e covariant \cite{II,IV} since 
$U_{\Lambda,y}^{-1}\underline{I}_{\vec p}U_{\Lambda,y}=
\underline{I}_{\overrightarrow{\Lambda^{-1} p}}$, 
$U_{\Lambda,y}^{-1}\underline{I}_{\vec k}U_{\Lambda,y}=
\underline{I}_{\overrightarrow{\Lambda^{-1} k}}$, which would not be possible if 
$\underline{I}_{\vec p}$ and $\underline{I}_{\vec k}$ were simply replaced in CAR and CCR by the functions 
$Z \chi(\vec p)$ and
$\tilde Z \tilde \chi(\vec k)$.
Our theory is nonlocal \cite{Efimov} but in an unusual sense.

The above representations have a status of toy models. The main message we have tried to convey is that reducibility of some type may be crucial for a consistent QFT, a viewpoint advocated also by Dirac in his last paper \cite{Dirac}. Generalizations to more abstract formalisms may be essential for the issue of gauge invariance, which is beyond our reach at the moment.  
In this context let us note the recent paper \cite{Jan} where a similar representation of CCR is discussed at the level of correlation-function approach to photons.

%\bigskip

My understanding of the problem was influenced by numerous discussions with Jan Naudts. I am grateful to UIA, Antwerp, for a financial support of this work.

\label{lastpage}


\begin{thebibliography}{99}
\small

\bibitem{I}
Czachor M, Non-canonical quantum optics, {\it J. Phys.} {\bf A33} 
(2000), 8081.

\bibitem{II}Czachor M and  Syty M, Non-canonical quantum optics (II): Poincar\'e covariant formalism and thermodynamic limit, quant-ph/0205011.

\bibitem{III}
Czachor M, States of light via reducible quantization, {\it Phys. Lett.} {\bf A313} (2003), 380.

\bibitem{IV}
Czachor M, Reducible field quantization (II): Electrons, quant-ph/0212061.

\bibitem{Efimov}Efimov G V, Nonlocal interactions of quantized fields, Nauka, Moscow, 1977 (in Russian).

\bibitem{Dirac}Dirac P A M, The future of atomic physics, Int. J. Theor. Phys. {\bf 23}, 677 (1984).

\bibitem{Jan}Naudts J, Kuna M and De Roeck W, Photon fields in a fluctuating spacetime, hep-th/0210188.

\end{thebibliography}
\end{document}